\documentclass[11pt]{article}
\usepackage{graphicx}
\usepackage{pgfplots}
\usepackage[colorlinks=true,citecolor=blue]{hyperref}
\usepackage{natbib}
\usepackage{graphicx} 
\usepackage{float}    
\usepackage{booktabs} 

\usepackage{caption}
\usepackage{amsfonts}
\usepackage{amsmath}
\usepackage{amssymb}
\usepackage{url}
\usepackage{fancyhdr}
\usepackage{indentfirst}
\usepackage{enumerate}
\usepackage{titlesec}
\usepackage{amsthm}
\usepackage{dsfont}
\usepackage[misc]{ifsym}
\usepackage{cases}
\usepackage{soul}      

\theoremstyle{definition}

\theoremstyle{plain}

\theoremstyle{remark}

\theoremstyle{definition}

\newcommand\E{\mathbb{E}}
\newcommand\R{\mathbb{R}}

\usepackage[onehalfspacing]{setspace}

\usepackage{bm}
\usepackage{tikz-qtree}
\usepackage{tikz}

\begin{document}
\title{An efficient method of posterior sampling for Poisson INGARCH models

\footnotetext{$\dag$Corresponding author. School of Mathematical Sciences, Nanjing Normal University, 1 Wenyuan Road, Nanjing 210023, China, E-mail: liuzw415@163.com}}
\author{Yixuan Fan$^1$, Zhengwei Liu$^{2\dag}$, and Fukang Zhu$^3$ \\[0cm]
{\small\it $^1$ Center for Applied Mathematics, KL-AAGDM, Tianjin University, Tianjin, 300072, China}\\
{\small\it $^2$ School of Mathematical Sciences, Nanjing Normal University, Nanjing, China}\\
{\small\it $^3$ School of Mathematics, Jilin University, Changchun, China}
}
\date{}

\maketitle
\begin{center}
\begin{minipage}{14.5truecm}
{\bf Abstract.}
We develop an efficient posterior sampling scheme for the Poisson INGARCH models. The proposed method is based on the approximation of the posterior density that exploits the Poisson limit of the negative binomial distribution. It allows us to rewrite the model in a form amenable to P\'{o}lya-Gamma data augmentation scheme, which yields simple conditionally Gaussian updates for the autoregressive coefficients. Sampling from the approximate posterior is straightforward via Gibbs-type iterations and remains numerically stable even under strong temporal dependence. Using this sampler as a proposal distribution will enhance the efficiency in Metropolis-Hastings algorithm and adaptive importance sampling. Numerical simulations indicate accurate posterior estimates, high effective sample sizes, and rapidly mixing chains.

{\bf Keywords:} 
Adaptive importance sampling; Pareto smoothing;
P\'{o}lya-Gamma data augmentation; 
Poisson INGARCH models 
\end{minipage}
\end{center}

\section{Introduction}
Integer-valued time series have been extensively studied in recent years, since they can be frequently encountered in the real world, such as certain diseases in epidemiology \citep{Chen2019}, criminal incidents \citep{Yang2023}, hospital operations \citep{Reboredo2023} and economics \citep{Jiang2024}. For modeling the dynamics of count time-series data, \cite{Ferland2006} first proposed the well-known Poisson integer-valued generalized autoregressive conditional heteroscedastic (INGARCH) model and established the theoretical framework. To capture both positive and negative dependence in count time series, \cite{Fokianos2011} introduced the log-linear Poisson autoregression under a nonlinear structure. A drawback, however, is that closed-form expressions for the unconditional mean and the autocorrelation function are typically unavailable. To remedy this limitation, \cite{Weiss2022} proposed the softplus Poisson INGARCH model, which keep the linear conditional mean and the ARMA-like autocorrelation. Following the notation of \cite{Sim2021}, the Poisson $\mbox{INGARCH}(p,q)$ model could be written as:
\begin{equation}
\left\{
\begin{aligned}
& X_t \mid \mathcal{F}_{t-1}\sim \mathrm{Poisson}(\lambda_t),\\
& \eta_t = \psi(\lambda_t),\\ 
& \eta_t = \alpha_0 + \sum_{i=1}^p \alpha_i\eta_{t-i} + \sum_{j=1}^q \beta_j\Upsilon(X_{t-j}),
\end{aligned}
\right.
\end{equation}
where $\mathcal F_{t-1} = \sigma\{X_s:s\le t-1\}$ is the past information set, $\psi(\cdot)$ is the response function and $\Upsilon(\cdot)$ is a transformation of past counts. Typical choices include the identity link in \cite{Fokianos2009}, the log-linear specification in \cite{Fokianos2011} and the softplus function in \cite{Weiss2022}.

Although maximum likelihood estimation is available for Poisson INGARCH models, Bayesian inference remains attractive because it provides a natural framework for uncertainty quantification and prediction under stationarity constraints. In particular, it allows inference not only on model parameters, but also on persistence measures, predictive distributions, and other nonlinear functionals. The main difficulty, however, is not the Bayesian formulation itself, but the lack of an efficient posterior computation method for Poisson INGARCH models. Existing Bayesian work on integer-valued time series has demonstrated the practical value of posterior inference inference in a variety of settings. For example, \cite{Xu2020} proposed an adaptive log-linear zero-inflated generalized Poisson INGARCH model with locally adaptive windows; the time-varying parameters are estimated via adaptive Bayesian MCMC, enabling posterior tracking of nonstationary crime counts with overdispersion and excess zeros. \cite{Yang2022} developed a new MCMC algorithm for self-exciting integer-valued threshold time series models by introducing latent variables. \cite{Chen2023} advanced Bayesian spatial count time series modeling by combining multivariate log-linear INGARCH dynamics and the continuous 
spatial structure, thereby unifying parameter inference, prediction and model selection. \cite{Chu2023} developed a log-linear Beta-negative binomial INGARCH model and adopted a Bayesian framework to quantify uncertainty in the unknown parameters. However, the Bayesian procedures in these works generally rely on a fixed proposal within MCMC. For the Poisson INGARCH models, it can be problematic when the posterior geometry varies substantially across the parameter space, especially under strong persistence or nonlinear link functions. As a result, posterior simulation may suffer from slow mixing and low computational efficiency.

A natural question is therefore how to construct a proposal mechanism that is both computationally efficient and adaptive to the local shape of the posterior. 
To address this gap, we adapt the ideas from the generalized linear mixed models (GLMMs) literature, where iterative weighted least squares (IWLS) and associated Gaussian approximations are routinely used to obtain accurate local quadratic expansions. \cite{Breslow1993} made a foundational contribution to practical inference for GLMMs by formalizing two closely related approximation frameworks, penalized and marginal quasi-likelihood. They justified it through a Laplace approximation to the integrated quasi-likelihood, thereby turning the otherwise intractable integration over the random effects into a computationally manageable estimation procedure.
\cite{Gamerman1997} then embedded the IWLS-based 
local Gaussian approximation within a Metropolis-Hastings (MH) framework and placing it inside a blockwise Gibbs sampler, yielding a practical and efficient MCMC algorithm for Bayesian inference in GLMMs. The key idea is to leverage second-order  
curvature information to construct high quality proposals while preserving exact posterior targeting through MH correction, and to accommodate extensions such as nested and heavy-tailed  structure of random effects.

A further challenge arises when the log-intensity is not linear in predictors, as in the softplus response function. In this case, the Gaussian proposal cannot be obtained directly from the exact recursive structure. To overcome this difficulty, we combine the state-dependent proposal with a gradient-informed local linearization of the log-intensity, which extends the same computational strategy beyond the log-linear specification. In this sense, the contribution of the paper is not merely to introduce another Bayesian estimator, but to provide a practical and extensible tool of posterior computation  for Poisson INGARCH models.
Related developments include \cite{Wenzel2019}, who derived a gradient-based approach to variational inference in Gaussian process classification under the logit link, and \cite{Glynn2019}, who proposed a Bayesian dynamic linear topic model that combines P\'{o}lya-Gamma data augmentation with a CLT-based Gaussian approximation to latent variables, substantially reducing the computational cost of MCMC. Importantly, because the MH acceptance probability is evaluated using the exact likelihood, the resulting Markov chain still targets the exact posterior, regardless of the approximation quality of the proposal; the approximation affects efficiency rather than correctness (see \citealp{Chib1995}; \citealp{Gamerman1997}). In the special case of an independent MH algorithm, the ideal proposal is the target posterior itself, which leads to acceptance probability one and independent and identically distributed draws. More generally, the efficiency of the independence sampler is governed by how closely the proposal density approximates the target (see \citealp{Mengersen1996}; \citealp{Keith2008}; \citealp{Lee2018}). Building on these ideas, we develop an efficient state-dependent proposal mechanism for a class of Poisson INGARCH models.

The main contributions of this paper are summarized as follows.
\begin{itemize}
\item 
We develop a state-dependent Gaussian proposal for Poisson INGARCH models by approximating the Poisson likelihood with the negative binomial likelihood and exploit the resulting P\'{o}lya-Gamma representation. Unlike the conventional fixed proposal distribution, our proposed mechanism adapts to the local posterior geometry. Importantly, the resulting proposal is calibrated through a MH correction based on the exact Poisson likelihood, so the approximation improves computational efficiency without changing the target posterior distribution. 

\item 
We extend this proposal framework beyond the log-linear specification to the softplus Poisson INGARCH model. Through the gradient-informed local linearization of the log-intensity, we enlarge the scope of the method and show that the proposed framework is not limited to models with linear recursions. This yields, to our knowledge, the first broadly applicable posterior computation tool for Poisson INGARCH models.

\item 
We formulate an adaptive importance sampling scheme that reuse the same proposal family while avoiding accept/reject steps. To address potential weight degeneracy, we further incorporate Pareto-smoothed adaptive importance sampling, together with the diagnostic to assess the reliability of the resulting estimates. This turns the proposal mechanism into a broader posterior computation framework rather than the MH sampler. 

\item 
Through the numerical simulation and empirical study, we demonstrate that the proposed methods deliver accurate posterior inference, stable computation, and favorable sampling efficiency across both log-linear and softplus Poisson INGARCH specifications.

\end{itemize}

The goal of this paper is to develop an efficient and broadly applicable Bayesian inference for a class of Poisson INGARCH models, under both log-linear and nonlinear specifications. Section \ref{sec2} introduces the key intuition behind the negative-binomial approximation for Poisson INGARCH likelihoods, together with a gradient-informed strategy for handling nonlinear log-intensity specifications. Section \ref{sec3} presents the estimation procedure and the proposed posterior sampling method under stationarity. Section \ref{sec4} studies finite-sample performance under various Poisson INGARCH scenarios. Section \ref{sec5} provides a real-data application.

\newpage
\section{A unified framework of posterior sampling}\label{sec2}

In this section, we develop a unified framework of
posterior sampling based on the approximation of the likelihood function for Poisson INGARCH models. The latent variables are integrated out and replaced by the plug-in conditional mean, yielding a state-dependent Gaussian proposal. We further extend the construction to the nonlinear response function 
through a gradient-informed Taylor linearization.

\subsection{A state-dependent proposal}
Let $\textbf{\textit{X}}=(X_0,X_1,\cdots,X_n)^\top$ be the counts of time series and $\pmb{\theta} = (\alpha_0,\alpha_1,\beta_1)^\top$ be the vector of autoregressive parameters, with the initial intensity $\lambda_0$ assumed known. For the class of Poisson INGARCH models, the probability mass function is given by $f(x_t\mid \mathcal{F}_{t-1}) = e^{-\lambda_t}\lambda_t^{x_t}/x_t!$ and the log-likelihood function is proportional to $\log L(\pmb{\theta}\mid \textit{\textbf{X}})\propto \sum_{t=1}^{n}x_t\log(\lambda_t)-\lambda_t$.
Motivated by the Poisson limit of the negative binomial distribution, we approximate it by:
\begin{align}
\tilde{f}(x_t\mid \mathcal{F}_{t-1})
= \binom{r_t + x_t -1}{r_t-1}p_t^{r_t}(1-p_t)^{x_t}.
\end{align}
It corresponds to the negative binomial distribution with overdispersion parameter $r_t$ and success probability $p_t = \frac{r_t}{\lambda_t + r_t}$. As $r_t\rightarrow\infty$, $\tilde{f}(x_t\mid \mathcal{F}_{t-1})$ converges to the Poisson distribution. 
Therefore, the Poisson INGARCH models have the approximate likelihood function as follows:
\begin{align}\label{approximate_likelihood}
\tilde{L}(\textbf{\textit{X}}\mid \pmb{\theta}) 
&=\prod_{t=1}^{n}\tilde{f}(x_t\mid \mathcal{F}_{t-1})\propto \prod_{t=1}^{n} 
\left(\frac{r_t}{r_t + \lambda_t}\right)^{r_t}\left(\frac{\lambda_t}{r_t +\lambda_t}\right)^{x_t}.
\end{align}

The identity established by \cite{Polson2013} forms the kernel of data augmentation scheme, that is, 
\begin{align}\label{PG_identity}
\frac{(e^{\psi})^a}{(1+e^{\psi})^b}
=2^{-b} e^{\kappa\psi} \int_{0}^{\infty}
e^{-\omega \psi^2/2}
f_{\mathrm{PG}}(\omega \mid b, 0)\;d\omega,\ \kappa=a-b/2.
\end{align}
It implies that the binomial or negative binomial likelihood can be expressed as a Gaussian scale mixture with respect to P\'{o}lya-Gamma distribution $\mathrm{PG}(b,0)$. Here $f_{\mathrm{PG}}(\omega\mid b,0)$ denotes the probability density function of $\mathrm{PG}(b,0)$. Moreover, the P\'{o}lya-Gamma family of distributions $\mathrm{PG}(b,c)$ can be obtained through an exponential tilting of $\mathrm{PG}(b,0)$ distribution,
\begin{align}\label{pdf_of_PG(b,c)}
f_{\mathrm{PG}}(\omega\mid b,c)
=\frac{\exp\left(-\frac{1}{2}\omega c^2\right)f_{\mathrm{PG}}(\omega\mid b,0)}{\int_{0}^{\infty}\exp\left(-\frac{1}{2}\omega c^2\right)f_{\mathrm{PG}}(\omega\mid b,0)\,d\omega}.
\end{align}
The approximate likelihood function is proportional to
\begin{align}
\tilde{L}(\textbf{\textit{X}}\mid \pmb{\theta}) 
& \propto \prod_{t=1}^{n} e^{\kappa_t\psi_t}\int_{0}^{\infty}e^{-\omega\psi_t^2/2}f_{\mathrm{PG}}(\omega\mid x_t+r_t,0)\,d\omega,
\end{align} 
where $\psi_t = \log(\lambda_t)-\log(r_t)$ and $\kappa_t = (x_t-r_t)/2$. Through introducing the auxiliary variables $\omega_t\mid\pmb{\theta}\sim \mathrm{PG}\big(r_t+x_t,\psi_t\big)$, the joint complete posterior distribution of $\pmb{\theta}$ becomes
\begin{align}\label{approximate_posterior}
p(\pmb{\theta}\mid \textbf{\textit{X}},\pmb{\omega})
&\propto \prod_{t=1}^{n}\exp(\kappa_t\psi_t)\int_{0}^{\infty} \exp(-\omega\psi_t^2/2)f_{\mathrm{PG}}(\omega\mid x_t+r_t,0)\,d\omega  \nonumber\\
&\quad \times \prod_{t=1}^{n}\frac{\exp\left(-\frac{1}{2}\omega\psi_t^2\right)f_{\mathrm{PG}}(\omega\mid x_t+r_t,0)}{\int_{0}^{\infty}\exp\left(-\frac{1}{2}\omega \psi_t^2\right)f_{\mathrm{PG}}(\omega\mid x_t+r_t,0)\,d\omega}\times\pi(\pmb{\theta}) \nonumber\\
&\propto \pi(\pmb{\theta})\prod_{t=1}^{n}\exp\left(\kappa_t\psi_t-\frac{1}{2}\omega\psi_t^2\right).
\end{align}

When the log-intensity is linear in predictors, i.e., $\log(\lambda_t) = \pmb{\theta}^\top\bm{d}_t$, 
where $\bm{d}_t$ denotes the time-varying design vector collecting the intercept and lagged predictors, the conditional posterior distribution is given by
\begin{align}\label{conditional_Gauss}
p(\pmb{\theta}\mid \textbf{\textit{X}},\pmb{\omega})
\propto \pi(\pmb{\theta})\exp\left(-\frac{1}{2}(\textbf{\textit{D}}\pmb{\theta}- \textbf{\textit{Z}})^\top\pmb{\Omega}(\textbf{\textit{D}}\pmb{\theta}- \textbf{\textit{Z}})\right),
\end{align}
with working response $\bm{Z} = \left(\frac{\kappa_1}{\omega_1},\cdots,\frac{\kappa_n}{\omega_n}\right)^\top$, design matrix $\bm{D} = (\bm{d}_1^\top,\cdots,\bm{d}_n^\top)$, diagonal covariance matrix $\pmb{\Omega} = \mathrm{diag}\{\omega_1,\cdots,\omega_n\}$ and $\pmb{\kappa}= (\kappa_1,\cdots,\kappa_n)^\top$. 
Given the Gaussian prior $\pi(\pmb{\theta})\sim N(\bm{b},\bm{B})$, the Gibbs sampler obtained from the approximate posterior distribution is
\begin{align}\label{Gibbs}
\left\{
\begin{aligned}
& \omega_t\mid \pmb{\theta} \sim \mathrm{PG}\big(r_t+x_t,\ \psi_t\big),\\ 
& \pmb{\theta}\mid \pmb{\omega},\textbf{\textit{X}}\sim N(\pmb{\mu}_{\omega},\textbf{V}_{\omega}),
\end{aligned}
\right.
\end{align}
where $\textbf{\textit{V}}_{\omega} = (\textbf{\textit{D}}^\top\pmb{\Omega}\textbf{\textit{D}} + \textbf{\textit{B}}^{-1})^{-1}$, $\pmb{\mu}_{\omega} = \textbf{\textit{V}}_{\omega}(\textbf{\textit{B}}^{-1}\textbf{\textit{b}} + \textbf{\textit{D}}^\top\pmb{\kappa})$.

However, large values of $r_t$ substantially increase the computational burden, as sampling P\'{o}lya-Gamma random variables with large shape parameters is computationally intensive. A draw from $\mathrm{PG}(b,c)$ is obtained as the sum of $b$ independent samples from $\mathrm{PG}(1,c)$. Each variate is generated via an accept-reject step with a proposal given by a mixture of an inverse-Gaussian and an exponential distribution. The method is available in the R package BayesLogit \citep{Polson2019}. In practice, it implies that the computational cost of sampling $\mathrm{PG}(r_t,\cdot)$ grows roughly linearly in $r_t$, which will substantially slow down the overall MCMC algorithm. Inspired by \cite{Gamerman1997} and \cite{DAngelo2023}, a natural strategy is to marginalize the joint posterior distribution conditional on the current value. To reduce the computational burden, we integrate out the latent variables of the conditional Gaussian distribution in (\ref{conditional_Gauss}). The marginal posterior can be written as a generally intractable form,
\begin{align}
p(\pmb{\theta}\mid \textbf{\textit{X}})
\propto \int_{}^{} p(\pmb{\theta}\mid \pmb{\omega},\textbf{\textit{X}})f\left(\pmb{\omega}\mid \pmb{\theta},\textbf{\textit{X}}\right)\,d\pmb{\omega}
= \E_{\pmb{\omega}\sim f(\cdot| \pmb{\theta},\textbf{\textit{X}})}\left[p(\pmb{\theta}\mid \pmb{\omega},\textbf{\textit{X}})\right].
\end{align}
Thus we approximate it via a first-order delta method. Specifically, taking $\pmb{\omega}_0=\E(\pmb{\omega}\mid \pmb{\theta}, \textbf{\textit{X}})$ gives the plug-in approximation
\begin{align}
\E_{\pmb{\omega}\mid \pmb{\theta},\textbf{\textit{X}}}[p(\pmb{\theta}\mid \pmb{\omega},\textbf{\textit{X}})]
\approx 
p(\pmb{\theta}\mid \E(\pmb{\omega}\mid \pmb{\theta},\bm{X}),\bm{X}).
\end{align}
At iteration $k$, we evaluate the conditional mean at the current state $\pmb{\theta}^{(k-1)}$ via
\begin{align*}
\bar\omega_t^{(k-1)}:= 
\mathbb{E}\!\left[\omega_t\mid \pmb{\theta}^{(k-1)},\textbf{\textit{X}}\right]
= \frac{r_t+x_t}{2\,\psi_t^{(k-1)}}\tanh\!\left(\frac{\psi_t^{(k-1)}}{2}\right).
\end{align*}
Plugging $\bar{\pmb{\omega}}^{(k-1)} = \pmb{\omega}$ into the posterior distribution (\ref{Gibbs}) yields the state-dependent Gaussian proposal distribution, that is,
\begin{equation}\label{final_proposal}
g\left(\pmb{\theta}^* \mid \pmb{\theta}^{(k-1)}\right)
\sim N\left(\pmb{\mu}^{(k-1)},\textbf{\textit{V}}^{(k-1)}\right), 
\end{equation}
where
\begin{align*}
\textbf{\textit{V}}^{(k-1)}=\left(\textbf{\textit{D}}^\top\bar{\pmb{\Omega}}^{(k-1)}\textbf{\textit{D}}+\textbf{\textit{B}}^{-1}\right)^{-1},\ 
\pmb{\mu}^{(k-1)}=\textbf{\textit{V}}^{(k-1)}\left(\textbf{\textit{D}}^\top\bar{\pmb{\kappa}}^{(k-1)}+\textbf{\textit{B}}^{-1}\textbf{\textit{b}}\right).
\end{align*}
Here $\bar{\pmb{\Omega}}^{(k-1)}=\mathrm{diag}\!\left(\bar\omega_1^{(k-1)},\ldots,\bar\omega_n^{(k-1)}\right)$ and $\bar{\kappa}_t^{(k-1)}=\bar\omega_t^{(k-1)}\log r_t+\frac{x_t-r_t}{2}$.

\subsection{Linearization of log-intensity}
While the canonical log-link specification makes the log-intensity linear in the predictors, this property may fail under more robust response functions. A representative example is the softplus function (see \citealp{Weiss2022}), which ensures a positive intensity while growing only linearly for large arguments, thereby avoiding the explosive growth 
induced by the exponential link.

Consider the softplus INGARCH recursion, $\lambda_t = s_c(\alpha_0 + \alpha_1 \lambda_{t-1} + \beta_1 X_{t-1})$, where $s_c(x) = c\log (1 + \exp(x/c))$ is the softplus function. Define the log-intensity $z_t(\pmb{\theta}) = \log(\lambda_t)$ and let $\pmb{\theta}^{(k-1)}$ be the current iterate at iteration $k$. Since $z_t(\pmb{\theta}) = \log(\lambda_t)$ is nonlinear in $\pmb{\theta}$, we adopt a first-order Taylor expansion around $\pmb{\theta}^{(k-1)}$, that is,
\begin{align}
z_t(\pmb{\theta})
&\approx z_t\!\left(\pmb{\theta}^{(k-1)}\right)
+ \nabla z_t\!\left(\pmb{\theta}^{(k-1)}\right)^\top
\left(\pmb{\theta}-\pmb{\theta}^{(k-1)}\right) \nonumber\\
&= o_t^{(k-1)} + J_t^{(k-1)\top}\pmb{\theta},
\end{align}
where 
\begin{align}
o_t^{(k-1)} = z_t\left(\pmb{\theta}^{(k-1)}\right) - J_t^{(k-1)\top}\pmb{\theta}^{(k-1)},\ J_t^{(k-1)}= \nabla z_t
\left(\pmb{\theta}^{(k-1)}\right).
\end{align}
This linearization is exact under the log-link case and serves as a local approximation for nonlinear links such as softplus. By the chain rule, we have
\begin{align}
J_t(\pmb{\theta})
= \frac{\partial \log(\lambda_t)}{\partial \pmb{\theta}}
= \frac{1}{\lambda_t}\frac{\partial \lambda_t}{\partial \pmb{\theta}}
= \frac{s_c'(\eta_t)}{\lambda_t}\frac{\partial\eta_t}{\partial\pmb{\theta}},
\end{align}
with $\eta_t = \pmb{\theta}^\top\bm{d}_t$. The derivative
$\partial \eta_t/\partial \pmb{\theta}$ is state-dependent because $\eta_t$ contains
$\lambda_{t-1}$, which itself depends on $\pmb{\theta}$. To make the computation of $J_t$ more clear and easier, we denote that
\begin{align}
h_t = \frac{\partial \lambda_t}{\partial \boldsymbol{\theta}} = s_c'(\eta_t)\, g_t, 
\ g_t = \frac{\partial \eta_t}{\partial \boldsymbol{\theta}}
= (1,\lambda_{t-1},X_{t-1})^\top + \alpha_1 h_{t-1},
\end{align}
with initialization $h_0=\mathbf{0}$ if $\lambda_0 = 0$. Hence, $\{(\lambda_t,g_t,h_t,J_t)\}_{t=1}^n$ can be obtained via a single forward recursion in $t$, with computational cost $\mathcal{O}(np)$ for $p=\dim(\boldsymbol{\theta})$.

The values of the parameters $r_1, \cdots, r_n$ need to be tuned in the MH algorithm and the adaptive importance sampling. However, jointly tuning these parameters becomes impractical as $n$ grows large. A straightforward alternative is to assign them a common value, but this results in a small effective sample size for some chains in our simulations. Alternatively, we regulate the discrepancy between the Poisson and negative binomial likelihoods. Specifically, we adpot the upper bound on the relative error between their cumulative distribution functions in \cite{Teerapabolarn2012}. For each time point $t$, let $X\sim\mathrm{Poi}(\lambda_t)$ and $V\sim\mathrm{NB}(r_t,p_t)$. Define the time-varying discrepancy
\begin{align*}
d_t = \sup_{x_t\geq 0}\left|\frac{P(X\leq x_t)}{P(V\leq x_t)} - 1 \right| = 1 - e^{-\lambda_t}\left(1+\frac{\lambda_t}{r_t}\right)^{r_t}.
\end{align*}
Therefore, we impose a time-invariant tolerance level $d$ on the discrepancy between Poisson and negative binomial distribution uniformly over time. This uniform constraint automatically determines the entire collection $\{r_t\}_{t=1}^n$. Consequently, for each observation $X_t$, the resulting proposal density is guaranteed to stay within the same prescribed distance from the target posterior. In practice, at the beginning of iteration $k$,
we compute $r_1,\cdots,r_n$ based on the current state $\pmb{\theta}^{(k-1)}$, and then construct the proposal $g(\pmb{\theta}^*\mid \pmb{\theta}^{(k-1)})$ accordingly.

\section{Bayesian inference under stationarity}\label{sec3}

In this section, we present the Bayesian inference using our proposed Gaussian proposal and calibrating it with the exact Poisson likelihood under stationarity. 
We also formulate an adaptive importance sampling approach that reuse the proposed proposal, 
while correcting for adaptation through importance weights. To stabilize potential weight degeneracy, the largest weights are Pareto-smoothed and assessed by a tail-shape diagnostic.

\subsection{MH-within-Gibbs sampling}
Given the counts of time series $\textbf{\textit{X}}=(X_0,X_1,\cdots,X_n)^\top$, we aim to implement Bayesian inference for the unknown parameter vector $\pmb{\vartheta} = (\alpha_0,\alpha_1,\beta_1,\lambda_0)^\top$. The parameter groups include (i) $\alpha_0$; (ii) $\{\alpha_1,\beta_1\}$; (iii) $\lambda_0$ with the assumption that they are priori-independent. The MCMC method provides an efficient way to generate samples from the joint posterior distribution. By Bayes theorem, the posterior distribution for each group is
\begin{align}
p(\pmb{\vartheta}_j\mid \textbf{\textit{X}},\pmb{\vartheta}_{-j}) \propto L(\textbf{\textit{X}}\mid \pmb{\vartheta})\pi(\pmb{\vartheta}_j),
\end{align}
where $\pmb{\vartheta}_j$ denotes each parameter group, with $j= 1, 2, 3$; $\pi(\pmb{\vartheta}_j)$ is the prior density, and $\pmb{\vartheta}_{-j}$ denotes the vector of all parameters excluding the component $\pmb{\vartheta}_j$.
\begin{enumerate}[(1)]
\item For the intercept $\alpha_0$, we consider a normal prior $\pi(\alpha_0)$. Then the full conditional posterior distribution of $\alpha_0$ is
\begin{align*}
p(\alpha_0\mid\pmb{\vartheta}_{-1},\textbf{\textit{X}})\propto L(\textbf{\textit{X}}\mid \pmb{\vartheta})\pi(\alpha_0).
\end{align*}
\item For the  prior of the parameter block $\{\alpha_1,\beta_1\}$, we employ the constrained normal distribution. The full conditional posterior distribution of $\{\alpha_1,\beta_1\}$ is 
\begin{align*}
p(\alpha_1,\beta_1\mid \pmb{\vartheta}_{-2},\textbf{\textit{X}})
&\propto L(\textbf{\textit{X}}\mid \pmb{\vartheta})\pi(\alpha_1)\pi(\beta_1)\mathbb{I}(A),
\end{align*}
where $\mathbb{I}(\cdot)$ denotes the indicator function and $A$ is the set satisfies stationarity condition for the corresponding models.

\item For the initial intensity $\lambda_0$, we specify a Gamma prior $\pi(\lambda_0)$ to ensure that it is strictly positive. The full conditional posterior distribution of $\lambda_0$ is 
\begin{align*}
p(\lambda_0\mid \pmb{\vartheta}_{-3},\textbf{\textit{X}})
\propto L(\textbf{\textit{X}}\mid \pmb{\vartheta}) \pi(\lambda_0).
\end{align*}
\end{enumerate}

\noindent\textbf{The MCMC procedure is provided as follows}.
\begin{description}
\item[Step 1.] Set initial values for $\pmb{\vartheta}^{(0)} = \left(\alpha_0^{(0)},\alpha_1^{(0)},\beta_1^{(0)},\lambda_0^{(0)}\right)^\top$. 

\item[Step 2a.] Generate candidate values $\pmb{\theta}^*$ from the proposal density $g\left(\pmb{\theta}^*\mid\pmb{\theta}^{(k-1)}\right)$. 

\item[Step 2b.] Keep the candidate values if $\pmb{\theta}^*$ satisfies the stationarity condition. 
Accept it with the acceptance probability 
\begin{align*}
R(\pmb{\theta}^*,\pmb{\theta}^{(k-1)}) = \min\left\{1, 
\frac{L(\pmb{\theta}^*\mid\bm{X})\times g\left(\pmb{\theta}^{(k-1)}\mid \pmb{\theta}^*\right)}{L\left(\pmb{\theta}^{(k-1)}\mid\bm{X}\right)\times g\left(\pmb{\theta}^*\mid \pmb{\theta}^{(k-1)}\right)}\right\},
\end{align*}
and set $\pmb{\theta}^{(k)} = \pmb{\theta}^*$. Otherwise, stay at $\pmb{\theta}^{(k)} = \pmb{\theta}^{(k-1)}$. It is computed based on the ratio of the true likelihood function 
as a calibration and the ratio of proposal densities. The numerator is the proposal density evaluated at $\pmb{\theta}^*$ and the denominator $g(\pmb{\theta}^{(k-1)}\mid\pmb{\theta}^*)$ is the $N\!\left(\pmb{\mu}^*,\textbf{\textit{V}}^*\right)$ density evalued at $\pmb{\theta}^{(k-1)}$, where $\pmb{\mu}^*$ and $\textbf{\textit{V}}^*$ have the same expression as $\pmb{\mu}^{(k)}$ and $\textbf{\textit{V}}^{(k)}$ but depend on $\pmb{\mu}^*$ instead of $\pmb{\mu}^{(k-1)}$.

\item[Step 3.] Generate candidate value $\lambda_0^*$ from the proposal density $\mathrm{Gamma}(a_1,b_1)$ and accept or reject it using a random walk MH algorithm. 

\item[Step 4.] Go to the next iteration or stop if the chain has converged. 
\end{description}

\subsection{Pareto-smoothed adaptive importance sampling} 
An analogous construction can be formulated within the importance sampling framework. To approximate posterior expectations of the form $\E[h(\pmb{\theta})]=\int h(\pmb{\theta})p(\pmb{\theta}\mid\bm{X})\,d\pmb{\theta}$, we employ importance sampling (IS), which replaces direct simulation from the posterior distribution by sampling from the proposal. Let $g(\cdot)$ denote the proposal density.
In typical Bayesian settings, the posterior is available only up to a normalizing constant, thus we use the self-normalized IS,
\begin{align}
\widehat{\E[h(\pmb{\theta})]}
=\frac{1}{\sum_{s=1}^S r_s}\sum_{s=1}^S r_s\,h\left(\pmb{\theta}^{(s)}\right), \ r_s=\frac{p(\pmb{\theta}^{(s)}\mid\bm{X})}{g(\pmb{\theta}^{(s)})}.
\end{align}

Rather than employing a fixed proposal, we consider an adaptive proposal that evolves during sampling. Specifically, we use the state-dependent proposal in (\ref{final_proposal}), but unlike MH, we do not accept or reject samples to construct a Markov chain targeting the posterior. Let $\pmb{\theta}^c$ denote the current conditioning value. At iteration $s$, we draw a new value from $\pmb{\theta}^{(s)}\sim g(\pmb{\theta}^{(s)}\mid \pmb{\theta}^c)$, and update it only upon an uphill move in posterior density. 
The corresponding importance ratio is computed using the proposal actually used at iteration $s$, $\tilde{r}_s=\frac{p(\pmb{\theta}^{(s)}\mid\bm{X})}{g(\pmb{\theta}^{(s)}\mid \pmb{\theta}^{c})}.$
This adaptive mechanism steers the proposal toward high-density regions while preserving IS correction through the denominator.

A persistent practical challenge in IS is weight degeneracy: if the proposal under-covers regions of non-negligible posterior mass, the ratios $\{\tilde{r}_s\}$ can exhibit heavy tails, producing a few extreme weights that dominate the Monte Carlo average and inflate variance. To mitigate this instability, we apply Pareto-smoothed adaptive importance sampling (PSAIS), which stabilizes the largest ratios through an explicit tail model rather than hard truncation. Let \(r_{(1)}\le \cdots \le r_{(S)}\) be the ordered ratios of $\{\tilde{r}_s\}$. PSAIS identifies a right-tail subset of size and fits a generalized Pareto distribution (GPD) to the excesses
\(y_i=r_{(S-M+i)}-\hat{u}\), \(i=1,\ldots,M\),
yielding estimates \((\hat{k},\hat{\sigma})\).
The largest $M$ ratios are then replaced by the corresponding fitted GPD quantiles,
while the remaining ratios are left unchanged, i.e., $w_{(s)}=r_{(s)}$ for $s\le S-M$.
Finally, we renormalize the smoothed ratios and compute the PSAIS-adjusted estimator. The estimated GPD shape parameter $\hat{k}$ serves as a diagnostic of tail heaviness and, consequently, the reliability of the IS approximation: larger $\hat{k}$ indicates heavier tails and more fragile Monte Carlo behavior. Following standard practice, we flag potentially unstable estimates when $\hat{k} > \min\{1-1/\log_{10}(S),\,0.7\}$, in which case either a larger sample size or an improved proposal distribution may be required.

\noindent\textbf{The PSAIS procedure is provided as follows}.
\begin{description}
\item[Step 1.] Set the initial adaptive center $\pmb{\theta}^{(0)}$, choose the Monte Carlo size $S$, and specify the target functional $h(\pmb{\theta})$.

\item[Step 2.]  For $s=1,\ldots,S$, draw
$\pmb{\theta}^{(s)} \sim g(\cdot\mid \pmb{\theta}^{c})$ and compute the raw importance ratio:
\begin{align*}
\tilde r_s=\frac{p(\pmb{\theta}^{(s)}\mid \bm{X})}{g(\pmb{\theta}^{(s)}\mid\pmb{\theta}^c)}.
\end{align*}
Update the center by the uphill rule:
\begin{align*}
\pmb{\theta}^{(s)}=
\begin{cases}
\pmb{\theta}^{(s)}, & \text{if}\ p(\pmb{\theta}^{(s)}\mid \bm{X}) > p(\pmb{\theta}^c \mid \bm{X}),\\
\pmb{\theta}^c, & \text{otherwise}.
\end{cases}
\end{align*}

\item[Step 3.] Sort $\{\tilde r_s\}_{s=1}^S$ increasingly as $\tilde r_{(1)}\le\cdots\le\tilde r_{(S)}$, choose
\begin{align*}
M=\left\lfloor\min(0.2 S,\,3\sqrt S)\right\rfloor,\ u=\tilde r_{(S-M)},
\end{align*}
fit a GPD to to the exceedances
$y_i=\tilde r_{(S-M+i)}-u,\ i=1,\ldots,M$.

\item[Step 4.] Replace the largest $M$ raw ratios by Pareto-smoothed values
\begin{align*}
w_{(S-M+z)}
=\min\!\left\{
F_{\mathrm{GPD}}^{-1}\!\left(\frac{z-1/2}{M}\right),
\max_{1\le j\le S}\tilde r_j
\right\},
\quad z=1,\ldots,M,
\end{align*}
and keep the remaining ratios unchanged $w_{(s)}=\tilde r_{(s)}$ for $s\le S-M$.

\item[Step 5.] Normalize weights and  compute the PSAIS estimator
\begin{align*}
\tilde w_s=\frac{w_s}{\sum_{j=1}^S w_j},\qquad
\widehat{\mathbb E}_{\text{PSAIS}}[h(\pmb{\theta})] 
=\sum_{s=1}^S \tilde w_s\,h(\pmb{\theta}^{(s)}).
\end{align*}
Use the estimated GPD shape parameter $\hat{k}$ as a stability diagnostic. If $\hat{k}$ is large (e.g., >0.7), increase $S$ or improve the proposal $g(\cdot\mid \pmb{\theta}^c)$.
\end{description}

\section{Numerical simulation}\label{sec4}
To assess the performance of the Bayesian estimator for Poisson INGARCH models, we present a simulation study under different scenarios with sample size  $n = 800$. The initial observation $X_0$ is given, we take the initial intensity $\lambda_0$ as an unknown parameter and employ the 
Bayesian approach for parameter estimation. We set the initial values $\pmb{\theta}^{(0)} =(0.1, 0.1, 0.1)^\top$ and $\lambda_0^{(0)}=1$. For each replication, we perform $N = 10000$ iterations by discarding $H = 5000$ iterations as a burn-in sample for inference in each scenario. All simulation results are based on 100 replications.

\subsection{Log-linear Poisson model}
As stated in \cite{Fokianos2011}, the log-link specification in the autoregressive dynamics that could accommodate both positive and negative dependence is defined as
\begin{align} \label{log-linear}
\nu_t = \alpha_0 + \alpha_1\nu_{t-1} + \beta_1\log(1+X_{t-1}),
\end{align}
which allows the observation to determine the direction of association. The stationary condition is $\alpha_0\in\R$ and
\begin{align}
&|\alpha_1| < 1,\ \beta_1 > 0,\ |\alpha_1+\beta_1| < 1,  \label{constraint_1}\\ 
& \mbox{or}\ |\alpha_1| < 1,\ \beta_1 < 0,\ |\alpha_1||\alpha_1+\beta_1| < 1. 
\end{align}
Unlike the identity link, it does not impose sign restrictions on the autoregressive coefficients. We adopt constrained normal priors $\pi(\pmb{\theta})$ defined by indicators $\mathbb{I}(A_j)$ for $j=1,2$. Building on the similarity in patterns observed across the dataset, we take assumption (\ref{constraint_1}) and the conditional expectation $\E\left[\omega_t\mid \pmb{\theta}^{(k-1)}\right] =\frac{r_t+x_t}{2\psi_t^{(k-1)}}\left(\frac{\exp(\textbf{\textit{d}}_t^\top\pmb{\theta}^{(k-1)}) - r_t}{\exp(\textbf{\textit{d}}_t^\top\pmb{\theta}^{(k-1)}) + r_t}\right)$ in the subsequent analysis. Without generality, we set the initial intensity $\lambda_0 = 3$ and consider the Gaussian prior $\pi(\pmb{\theta})\sim N(\bm{b},\bm{B})$, with  $\textbf{\textit{b}}=(0, 0, 0)^\top,\ \textbf{\textit{B}}=\mbox{diag}\{1, 1, 1\}$.  
\begin{description}
\item[Scenario A1.] $\pmb{\theta} = (0.3, 0.2, 0.6)^\top$. 

\item[Scenario A2.] $\pmb{\theta} = (0.2, 0.3, 0.4)^\top$.
\end{description}

\begin{figure}[t]
\centering
\includegraphics{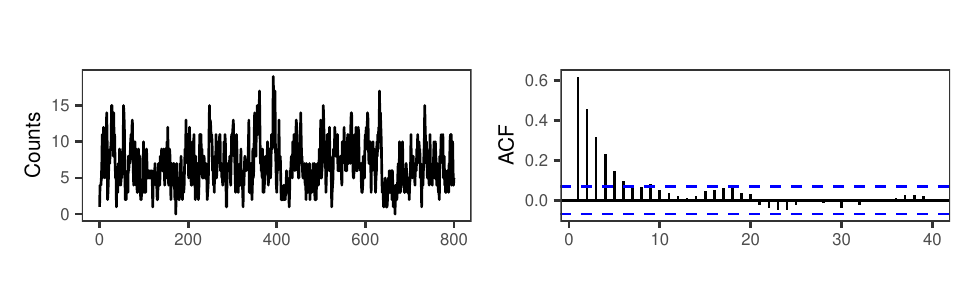}
\caption{Sample path and acf of Scenario A1.}
\label{A1}
\includegraphics{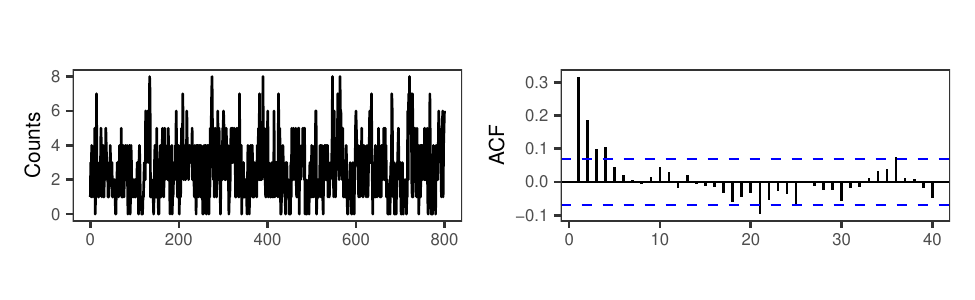}
\caption{Sample path and acf of Scenario A2.}
\label{A2}
\end{figure}

\begin{figure}[t]
\centering
\includegraphics[width=0.9\linewidth]{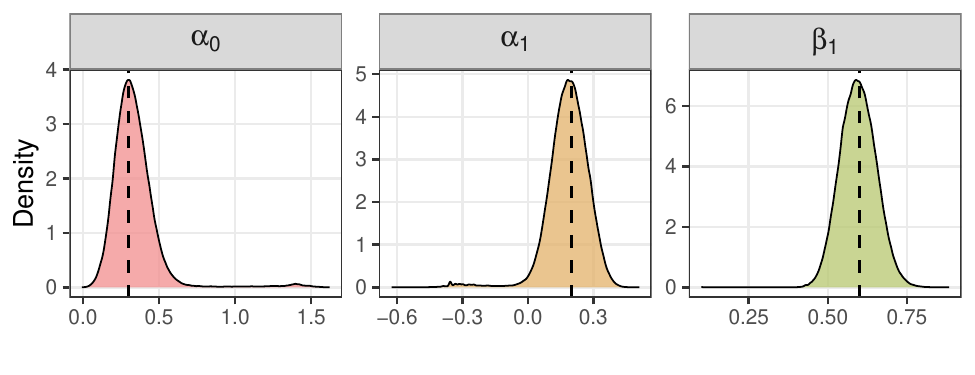}
\caption{Posterior distribution of the samples fitted to the simulated dataset illustrated in Fig.~\ref{A1}.}
\label{fig:posterior_A1}
\end{figure}

\begin{figure}[t]
\centering
\includegraphics[width=0.9\linewidth]{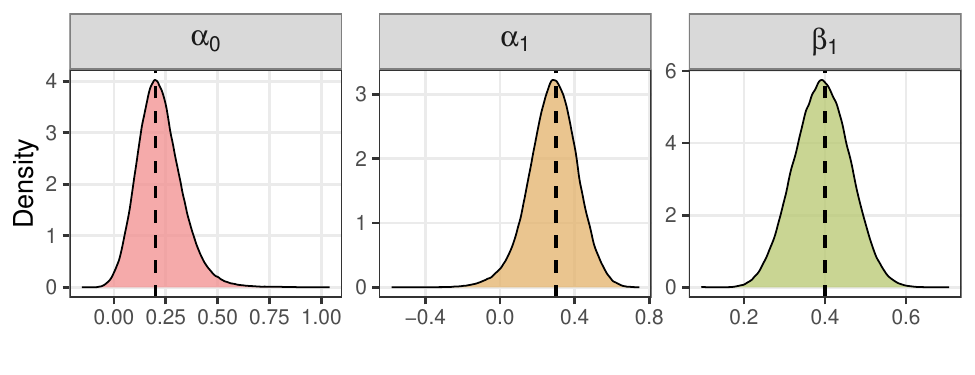}
\caption{Posterior distribution of the samples fitted to the simulated dataset illustrated in Fig.~\ref{A2}.}
\label{fig:posterior_A2}
\end{figure}

\begin{table}[H]
\centering
\caption{Parameter estimation accuracy for Scenario A1.}
\label{table:A1}
\begin{tabular}{ccccc@{\hspace{1em}}ccc@{\hspace{1em}}ccc}
\toprule
 &  & \multicolumn{3}{c}{MLE} & \multicolumn{3}{c}{Efficient MH} & \multicolumn{3}{c}{PSAIS} \\
\cmidrule(lr){3-5}\cmidrule(lr){6-8}\cmidrule(lr){9-11}
Param & True & Estimate & RMSE & MAD & Estimate & RMSE & MAD & Estimate & RMSE & MAD \\
\midrule
$\alpha_0$ & 0.3 & 0.323 & 0.0818 & 0.0650 & 0.321 & 0.0762 & 0.0607 & 0.312 & 0.0671 & 0.0529 \\
$\alpha_1$ & 0.2 & 0.195 & 0.0597 & 0.0461 & 0.192 & 0.0557 & 0.0469 & 0.199 & 0.0504 & 0.0394 \\
$\beta_1$  & 0.6 & 0.593 & 0.0397 & 0.0297 & 0.597 & 0.0399 & 0.0334 & 0.595 & 0.0388 & 0.0310 \\
\bottomrule
\end{tabular}
\end{table}

\begin{table}[H]
\centering
\caption{Parameter estimation accuracy for Scenario A2.}
\label{table:A2}
\begin{tabular}{ccccc@{\hspace{1em}}ccc@{\hspace{1em}}ccc}
\toprule
 &  & \multicolumn{3}{c}{MLE} & \multicolumn{3}{c}{Efficient MH} & \multicolumn{3}{c}{PSAIS} \\
\cmidrule(lr){3-5}\cmidrule(lr){6-8}\cmidrule(lr){9-11}
Param & True & Estimate & RMSE & MAD & Estimate & RMSE & MAD & Estimate & RMSE & MAD \\
\midrule
$\alpha_0$ & 0.2 & 0.219 & 0.0852 & 0.0677 & 0.222 & 0.0735 & 0.0569 & 0.229 & 0.0784 & 0.0610 \\
$\alpha_1$ & 0.3 & 0.297 & 0.1005 & 0.0818 & 0.284 & 0.0905 & 0.0719 & 0.275 & 0.0853 & 0.0679 \\
$\beta_1$  & 0.4 & 0.391 & 0.0486 & 0.0386 & 0.393 & 0.0515 & 0.0419 & 0.396 & 0.0407 & 0.0335 \\
\bottomrule
\end{tabular}
\end{table}

\subsection{Softplus Poisson model}
For softplus Poisson INGARCH model in \cite{Weiss2022}, the recursive structure is defined as,
\begin{align}\label{softplus}
\lambda_t = s_c(\alpha_0 + \alpha_1 \lambda_{t-1} + \beta_1 X_{t-1}),
\end{align}
where softplus function $s_c(x) = c\log (1 + \exp(x/c))$ serves as the response function. The set statisfies the stationary condition is, 
$$
A=\{\alpha_0>0,\alpha_1,\beta_1\geq 0, \alpha_1+\beta_1 < 1\}.
$$ 
We consider the Gaussian prior $\pi(\pmb{\theta})\sim N(\bm{b},\bm{B})$, with  $\textbf{\textit{b}}=(0, 0, 0)^\top,\ \textbf{\textit{B}}=\mbox{diag}\{0.15^2, 0.15^2, 0.15^2\}$. 
\begin{description}
\item[Scenario B1.] $\pmb{\theta} = (0.3, 0.4, 0.25)^\top$. 


\item[Scenario B3.] $\pmb{\theta} = (0.25, 0.35, 0.4)^\top$. 
\end{description}

\begin{figure}[t]
\centering
\includegraphics{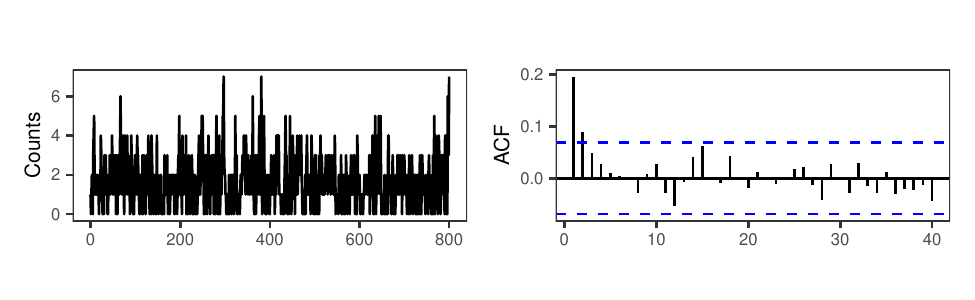}
\caption{Sample path and acf of Scenario B1.}
\label{B1}
\includegraphics{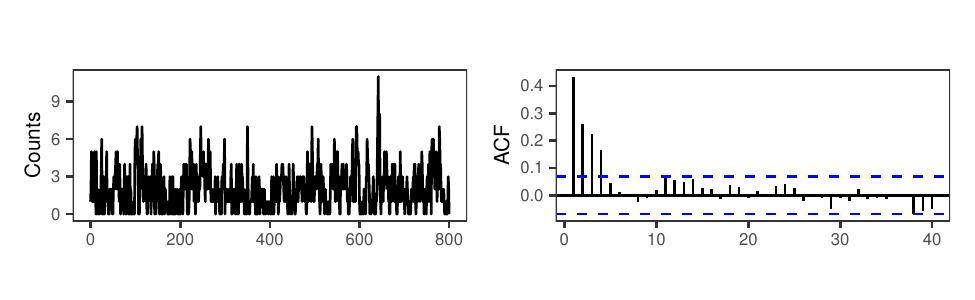}
\caption{Sample path and acf of Scenario B3.}
\label{B3}
\end{figure}

\begin{figure}[t]
\centering
\includegraphics{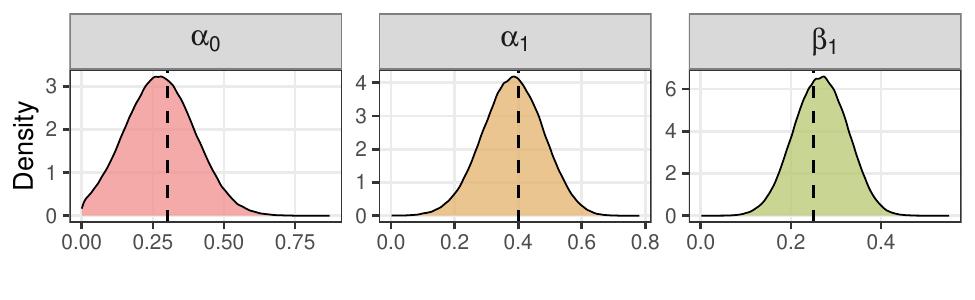}
\caption{Posterior distribution of the samples fitted to the simulated dataset illustrated in Fig. \ref{B1}.}
\label{fig:posterior_B1}
\end{figure}

\begin{figure}[t]
\centering
\includegraphics{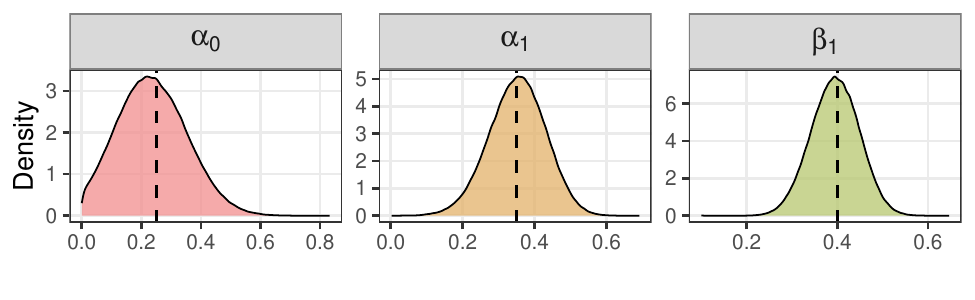}
\caption{Posterior distribution of the samples fitted to the simulated dataset illustrated in Fig. \ref{B3}.}
\label{fig:posterior_B3}
\end{figure}

\begin{table}[H] 
\centering
\caption{Parameter estimation accuracy for Scenario B1.}
\label{table:B1}
\begin{tabular}{ccccc@{\hspace{1em}}ccc@{\hspace{1em}}ccc}
\toprule
 &  & \multicolumn{3}{c}{MLE} & \multicolumn{3}{c}{Efficient MH} & \multicolumn{3}{c}{PSAIS} \\
\cmidrule(lr){3-5}\cmidrule(lr){6-8}\cmidrule(lr){9-11}
Param & True & Estimate & RMSE & MAD & Estimate & RMSE & MAD & Estimate & RMSE & MAD \\
\midrule
$\alpha_0$ & 0.30 & \textbf{0.360} & 0.2281 & 0.1810 & 0.277 & 0.0575 & 0.0442 & 0.296 & 0.0454 & 0.0355 \\
$\alpha_1$ & 0.40 & \textbf{0.366} & 0.1608 & 0.1274 & 0.383 & 0.0523 & 0.0423 & 0.382 & 0.0530 & 0.0435 \\
$\beta_1$  & 0.25 & 0.247 & 0.0488 & 0.0401 & 0.266 & 0.0451 & 0.0376 & 0.259 & 0.0340 & 0.0268 \\
\bottomrule
\end{tabular}
\end{table}

\begin{table}[H]
\centering
\caption{Parameter estimation accuracy for Scenario B3.}
\label{table:B3}
\begin{tabular}{ccccc@{\hspace{1em}}ccc@{\hspace{1em}}ccc}
\toprule
 &  & \multicolumn{3}{c}{MLE} & \multicolumn{3}{c}{Efficient MH} & \multicolumn{3}{c}{PSAIS} \\
\cmidrule(lr){3-5}\cmidrule(lr){6-8}\cmidrule(lr){9-11}
Param & True & Estimate & RMSE & MAD & Estimate & RMSE & MAD & Estimate & RMSE & MAD \\
\midrule
$\alpha_0$ & 0.25 & \textbf{0.309} & 0.1828 & 0.1411 & 0.239 & 0.0653 & 0.0541 & 0.246 & 0.0540 & 0.0427 \\
$\alpha_1$ & 0.35 & 0.330 & 0.1087 & 0.0860 & 0.351 & 0.0429 & 0.0343 & 0.349 & 0.0509 & 0.0420 \\
$\beta_1$  & 0.4 & 0.391 & 0.0428 & 0.0352 & 0.395 & 0.0356 & 0.0299 & 0.396 & 0.0345 & 0.0280 \\
\bottomrule
\end{tabular}
\end{table}

\newpage

As noted by \cite{Fokianos2009}, in Poisson INGARCH models, the maximum likelihood estimator (MLE) of the intercept parameter is often less robust than the other dynamic parameters. In particular, under highly persistent or near-nonstationary dynamics, the intercept parameter governing the long-run level 
tends to be weakly identified. Its estimator may exhibit unsatisfactory bias and variability in the simulation result. \cite{Pei2024}  addressed it for negative binomial INGARCH by estimating the intercept via a marginal likelihood approach. Building on these insights, we conduct numerical simulations to compare the standard maximum likelihood estimator with our proposed estimators for both log-linear and softplus Poisson INGARCH models.

The convergence diagnostics are reported via the trace plot and posterior mean for the unknown parameters across all simulation scenarios in the Appendix. For each parameter, the trace plot shows stable exploration of the posterior support without visible long-term drift, while the cumulative posterior mean stabilizes rapidly after the burn-in period, indicating satisfactory mixing and convergence of the sampler. The marginal posterior distributions of the autoregressive parameters under the proposed Bayesian estimation procedure are summarized in Figs.~\ref{fig:posterior_A1}-\ref{fig:posterior_A2} and \ref{fig:posterior_B1}-\ref{fig:posterior_B3}. Each panel displays the posterior density based on the MCMC output; the vertical dashed line marks the true value used to generate the simulated series. For the log-linear specification, the posterior mass is concentrated around the true parameter values, indicating that the parameters are well identified. The posteriors of $\alpha_1$ and $\beta_1$ reflect strong information in the data regarding short-run feedback and persistence. The posterior of $\alpha_0$ is also centered near the true value, with slightly heavier tails, suggesting comparatively greater uncertainty in the baseline level than in the dependence parameters. For the softplus response function, the figures exhibit similarly well-behaved posteriors that remain centered at the true values across both datasets. Relative to $\alpha_0$ and $\alpha_1$, the persistence parameter $\beta_1$ is more sharply identified, most notably in Scenario B2. It highlights that the dependence structure is strongly informed by the observed temporal dynamics. Overall, these posterior summaries suggest good finite-sample recovery of the parameters for both the log-linear and softplus formulations.

Tables~\ref{table:A1}-\ref{table:B3} compare the finite-sample parameter-estimation accuracy of MLE, Efficient MH and PSAIS for two class of INGARCH models across repeated simulations. Here root mean squared error (RMSE) penalizes large estimation errors more heavily via squaring, whereas mean absolute deviation (MAD) summarizes typical absolute error and is more robust to occasional outliers. Overall, all three methods recover the true parameters reasonably well, whereas the two Bayesian estimators generally achieve smaller RMSE and MAD than MLE, indicating improved estimation accuracy and robustness in finite samples.

The advantage of the proposed Bayesian procedures is more pronounced under the softplus specification. In Scenarios B1 and B3, the MLE exhibits substantially larger bias and variability for the intercept parameter; for example, in Scenario B1 its estimate is 0.360 for a true value of 0.30, with RMSE 0.2281, whereas the corresponding RMSE values are reduced to 0.0575 for Efficient MH and 0.0454 for PSAIS. A similar pattern is observed in Scenario B3. By contrast, the persistence parameter is estimated more accurately across all methods, although PSAIS still tends to deliver the smallest RMSE and MAD. These results suggest that the proposed Bayesian methods, particularly PSAIS, provide more accurate and stable inference than MLE, with especially notable gains for the intercept parameter.

\section{Real example}\label{sec5}
\begin{figure}[t]
\centering
\includegraphics{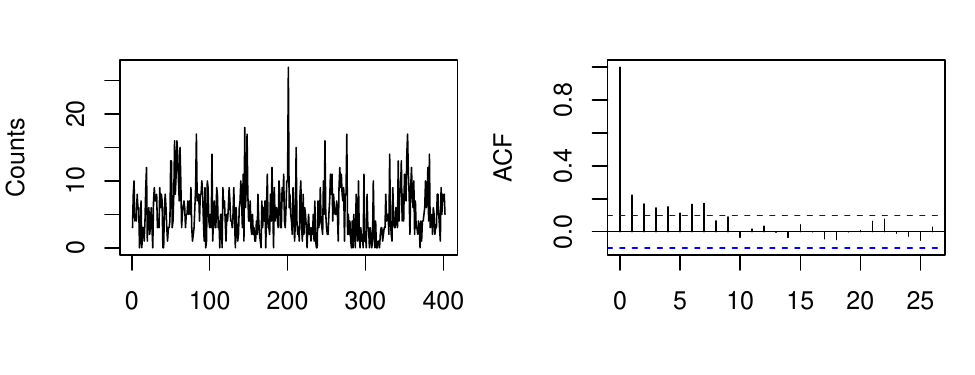}
\caption{Sample path and ACF of the GLT trade-count series (5-minute intervals).}
\label{fig:GLT_data}
\end{figure}

We analyze a univariate count time series $\{X_t\}_{t=1}^T$ for the P.H. Glatfelter Company (GLT) from the NYSE Trades and Quotes dataset. The observations record the number of trades within 5-minute intervals between 9{:}45 a.m. and 4{:}00 p.m. over the first quarter of 2005, and the same dataset was previously studied by \cite{Jung2011}. In this paper, we focus on the segment $\{X_t\}_{t=2500}^{2901}$ of the GLT
series, which contains $T=401$ five-minute observations. Since there are 75 five-minute intervals per trading day, this segment spans approximately 5 
trading days. The selected window retains pronounced short-lag dependence typical of high-frequency trading counts. Fig.~\ref{fig:GLT_data} displays the sample path and the sample ACF function, indicating strong serial dependence at short lags and motivating an INGARCH-type specification.

We fit the dataset with both log-linear and softplus Poisson INGARCHs.  For the softplus specification, we employ the gradient-informed local linearization of the log-intensity so that the same state-dependent Gaussian proposal mechanism applies. We assign a Gaussian prior $\pmb{\theta}\sim N(\bm{b},\bm{B})$ and restrict sampling to the stationarity region. Posterior inference is carried out using efficient Metropolis-Hastings with a state-dependent Gaussian proposal calibrated by the exact Poisson likelihood. We additionally report PSAIS estimates based on the same proposal family. The negative-binomial shape parameters $\{r_t\}$ used to construct the proposal are selected automatically by imposing an upper bound $d$ on the
Poisson-negative-binomial discrepancy, thereby controlling the approximation error over time.

\begin{figure}[t]
\centering
\includegraphics{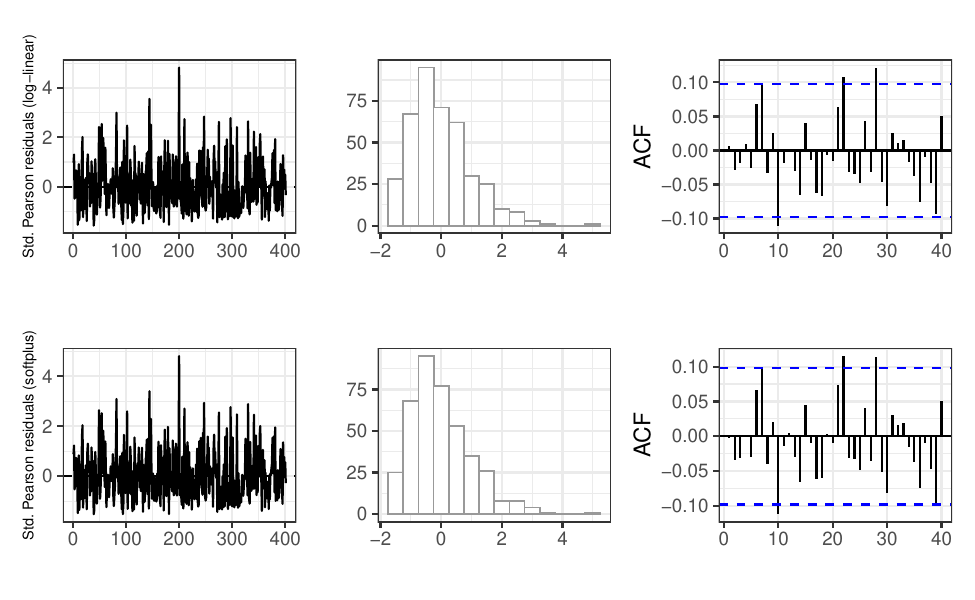}
\caption{Trajectory plot, histogram and ACF plot of the standardized residuals are obtained from the fitted models.}
\label{fig:GLT_residual_diagnostics}
\end{figure}

The marginal posterior densities, trace plots and running means are provides in the Appendix, indicating satisfactory mixing for both specifications. To assess model adequacy, we follow \cite{Jung2006} and compute standardized Pearson residuals as follows:
\begin{align}\label{eq:pearson_resid}
a_t = \frac{X_t-\mathbb{E}(X_t\mid\mathcal{F}_{t-1})}{\sqrt{\mathrm{Var}(X_t\mid\mathcal{F}_{t-1})}},
\end{align}
where $\mathbb{E}(X_t\mid\mathcal{F}_{t-1})$ and $\mathrm{Var}(X_t\mid\mathcal{F}_{t-1})$ are the model-implied conditional mean and variance. For a well-specified model, $\{a_t\}$ should fluctuate around zero and exhibit no significant serial correlation. Fig.~\ref{fig:GLT_residual_diagnostics} presents the trajectory plot, histogram and ACF of the standardized residuals for the log-linear and softplus models. The residual ACFs show no statistically significant autocorrelation, suggesting that both models adequately capture the short-range dependence in the GLT series.

We further compare the two models using one-step-ahead predictive performance. Let $\hat m_t=\mathbb{E}(X_t\mid\mathcal{F}_{t-1})$ denote the posterior predictive mean. We report the mean absolute error (MAE) and the root mean squared error (RMSE),
\begin{align}
\mathrm{MAE}=\frac{1}{T}\sum_{t=1}^T\bigl|X_t-\hat m_t\bigr|,
\qquad
\mathrm{RMSE}=\sqrt{\frac{1}{T}\sum_{t=1}^T\bigl(X_t-\hat m_t\bigr)^2}.
\end{align}
In addition, we evaluate density forecasts via the one-step-ahead log predictive density (LPD),
$\sum_{t}\log p(X_t\mid\mathcal{F}_{t-1})$, where $p(\cdot\mid\mathcal{F}_{t-1})$ is obtained by averaging the Poisson likelihood over posterior draws. The resulting metrics are summarized in Table~\ref{tab:fit_compare}. Overall, the two models deliver very similar point forecasts: the MAE values are virtually identical, and the RMSE values are close. In terms of density forecasting, the log-linear Poisson INGARCH attains a slightly higher one-step-ahead log predictive density compared with the softplus model, indicating a modest improvement in out-of-sample predictive likelihood. Taken together,
these results suggest that both link functions provide comparable predictive accuracy on the GLT series,
with a mild advantage for the log-linear specification in density forecasts.
\begin{table}[t]
\centering
\caption{Model fit and predictive performance for the GLT trade counts.}
\label{tab:fit_compare}
\begin{tabular}{lcc}
\hline
& Log-linear Poisson INGARCH & Softplus Poisson INGARCH \\
\hline
One-step-ahead LPD & -215.4767 & -217.6691 \\
MAE  & 2.6753  & 2.6735 \\
RMSE &  3.4379 &  3.4751 \\
\hline
\end{tabular}
\end{table}

\section{Conclusion}
In this paper, we develop an efficient and broadly applicable Bayesian posterior computation framework for Poisson INGARCH models. By approximating the Poisson likelihood through its negative binomial limit and exploiting the resulting P\'{o}lya-Gamma representation, we construct a state-dependent Gaussian proposal that 
preserves exact posterior targeting through MH correction. The same idea is further extended to the softplus specification via a gradient-informed local linearization. Numerical studies show that the proposed methods deliver accurate parameter recovery, stable computation, and favorable sampling efficiency across both log-linear and softplus Poisson INGARCH models, with PSAIS generally achieving the best overall estimation accuracy and robustness. The empirical analysis likewise suggests that both link functions provide comparable predictive performance, with the log-linear specification showing a slight advantage in density forecasting for the GLT trade-count series. Overall, the proposed framework offers a practical and extensible tool for Bayesian inference in Poisson INGARCH models and provides a useful foundation for future extensions to more general count time-series settings.

Although the present paper focuses on Poisson INGARCH models, the proposed posterior computation strategy may be extended to a broader class of non-negative integer-valued time-series models. In particular, when combined with a composite likelihood approach in \cite{Piancastelli2025}, it is also applicable to negative binomial INGARCH model  and potentially to other extensions of the Poisson framework considered in the literature (see \citealp{Xu2022}, \citealp{Xu2026}, and \citealp{Zhu2026}).

\section*{Acknowledgements}
\vspace{1mm}
Fan's work is supported by the National Natural Science Foundation of China (No. 12501378). Liu's work is supported by the Natural Science Foundation of Jiangsu Province (No. BK20240574). Zhu's work is supported by the National Natural Science Foundation of China (No. 12271206), and the Scientific Research Project Funding from the Education Department of Jilin Province (No. JJKH20261608KJ).

\clearpage
\section*{Appendix}
\vspace{3mm}

\begin{figure}[H]
\centering
\includegraphics{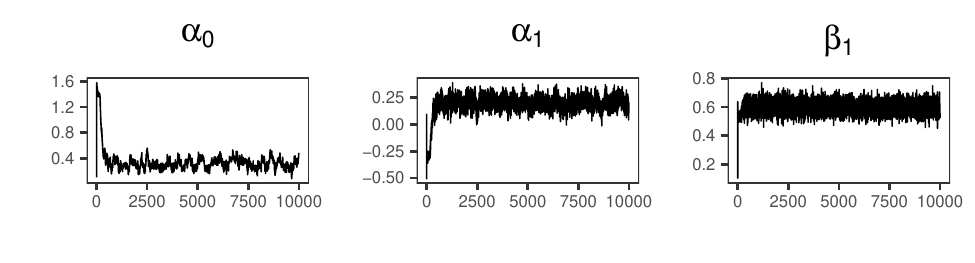}
\includegraphics{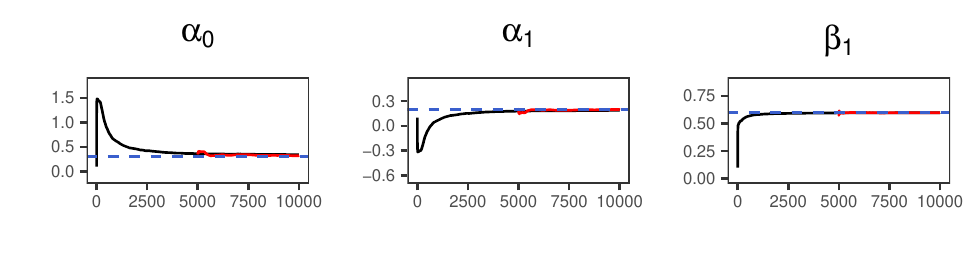}
\caption{Trace plot and posterior mean of Scenario A1.}
\end{figure}

\begin{figure}[H]
\centering
\includegraphics{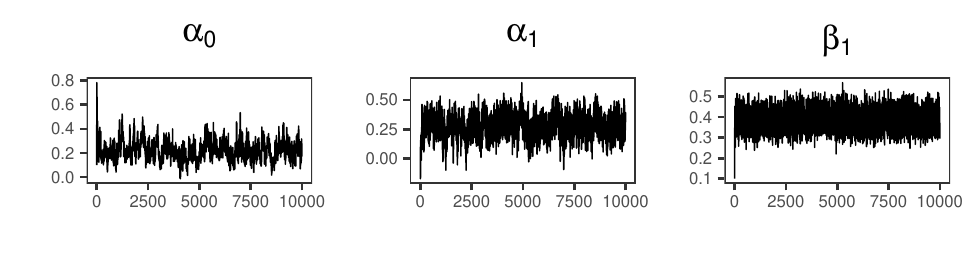}
\includegraphics{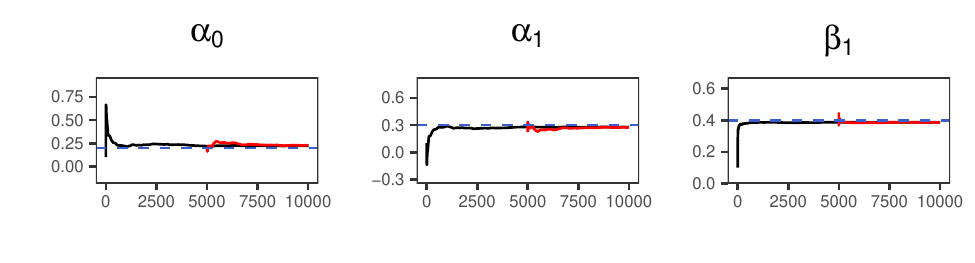}
\caption{Trace plot and posterior mean of Scenario A2.}
\end{figure}

\begin{figure}[H]
\centering
\includegraphics{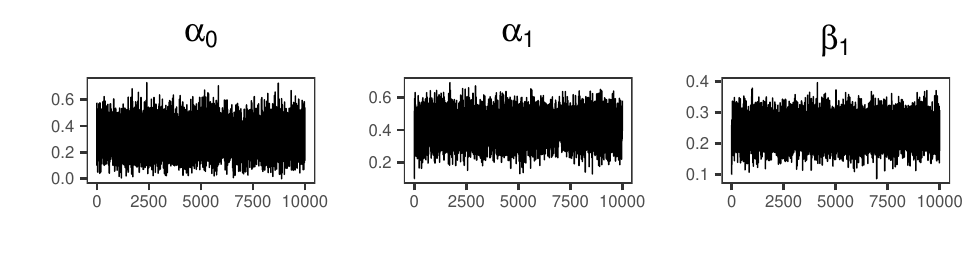}
\includegraphics{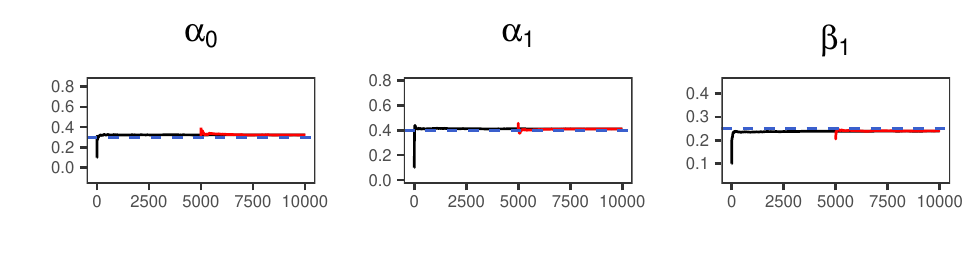}
\caption{Trace plot and posterior mean of Scenario B1.}
\end{figure}

\begin{figure}[H]
\centering
\includegraphics{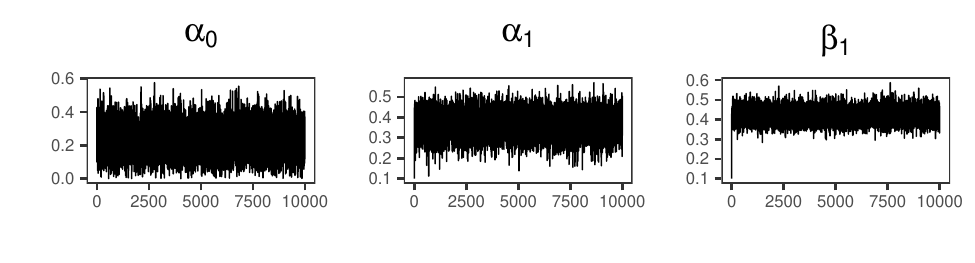}
\includegraphics{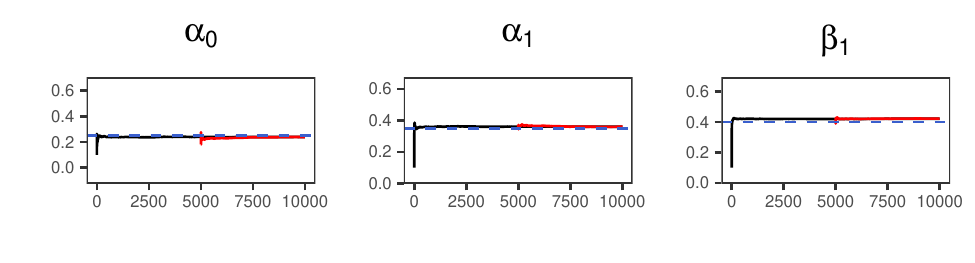}
\caption{Trace plot and posterior mean of Scenario B3.}
\end{figure}

\begin{figure}[H]
\centering
\includegraphics{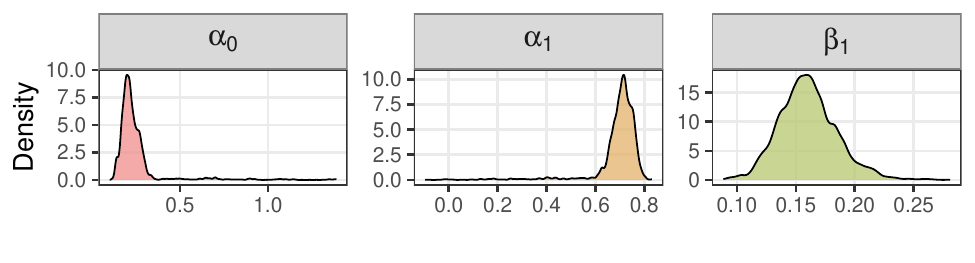}
\caption{The marginal posterior distribution of the trading days of the GLT data with log-linear model.}
\label{fig:GLT_ll_distribution}
\end{figure}

\begin{figure}[H]
\centering
\includegraphics{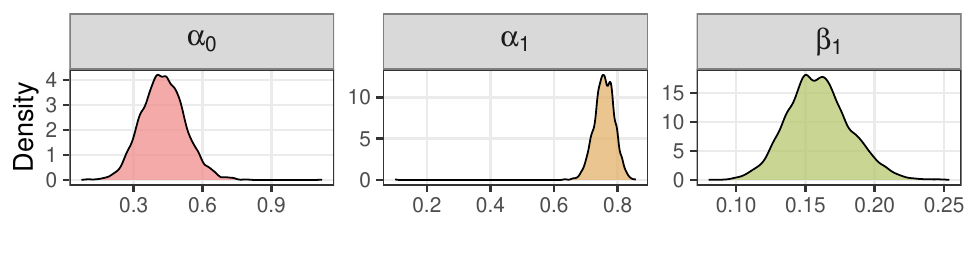}
\caption{The marginal posterior distribution of the trading days of the GLT data with softplus model.}
\label{fig:GLT_st_distribution}
\end{figure}

\begin{figure}[H]
\centering
\includegraphics{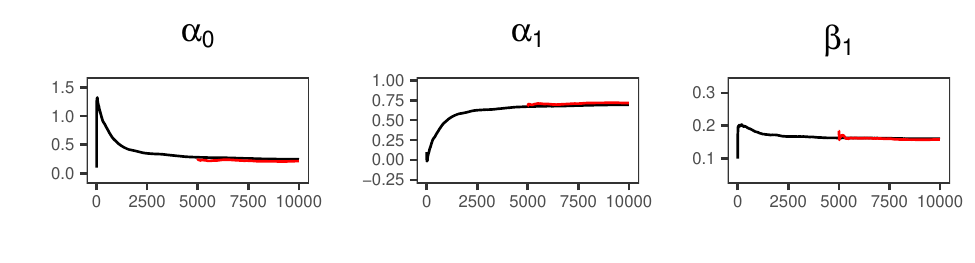}
\includegraphics{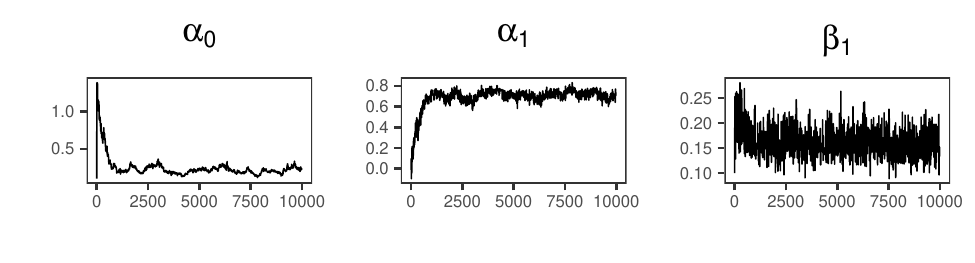}
\caption{The posterior mean and trace plot of the trading days of the GLT data with log-linear model.}
\label{fig:GLT_ll_posterior_trace}
\end{figure}

\begin{figure}[H]
\centering
\includegraphics{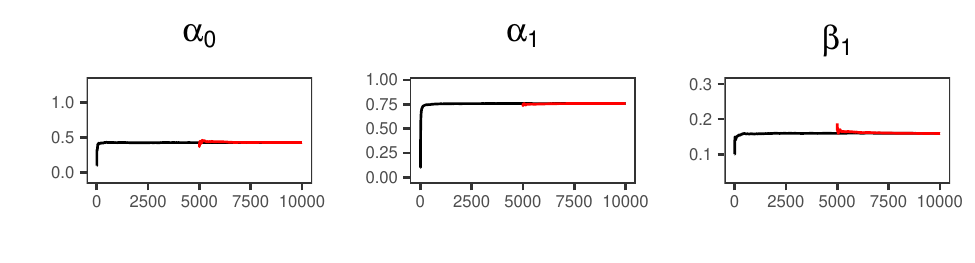}
\includegraphics{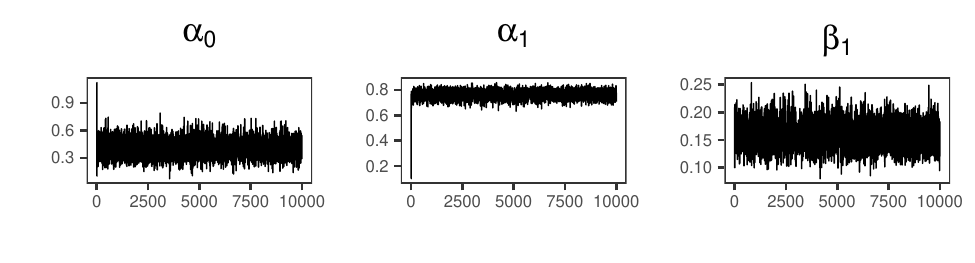}
\caption{The posterior mean and trace plot of the trading days of the GLT data with softplus model.}
\label{fig:GLT_st_posterior_trace}
\end{figure}


{
\small
\begin{thebibliography}{10}
\label{ref}
\bibitem[\protect\citeauthoryear{Breslow and
Clayton}{Breslow and Clayton}{1993}]{Breslow1993} 
Breslow, N. E., \& Clayton, D. G. (1993). Approximate inference in generalized linear mixed models. \textit{Journal of the American statistical Association}, \textbf{88}(421), 9-25.



\bibitem[\protect\citeauthoryear{Chen et al.}{Chen et al.}{2019}]{Chen2019} 
Chen, C. W., Khamthong, K., \& Lee, S. (2019). Markov switching integer-valued generalized auto-regressive conditional heteroscedastic models for dengue counts. \textit{Journal of the Royal Statistical Society Series}, 
\textbf{68}(4), 963-983.
\bibitem[\protect\citeauthoryear{Chen et al.}{Chen et al.}{2023}]{Chen2023}
Chen, C. W., Chen, C. S., \& Hsiung, M. H. (2023). Bayesian modeling of spatial integer-valued time series. \textit{Computational Statistics $\&$ Data Analysis}, \textbf{188}, 107-827.
\bibitem[\protect\citeauthoryear{Chib and Greenberg}{Chib and Greenberg}{1995}]{Chib1995} 
Chib, S., \& Greenberg, E. (1995). Understanding the Metropolis-Hastings algorithm. \textit{The American Statistician}, \textbf{49}(4), 327-335.

\bibitem[\protect\citeauthoryear{Chu and Yu}{Chu and Yu}{2023}]{Chu2023}
Chu, Y., \& Yu, K. (2024). Bayesian log-linear beta-negative binomial integer-valued Garch model. \textit{Computational Statistics}, \textbf{39}(3), 1183-1202.



\bibitem[\protect\citeauthoryear{D'Angelo and Canale}{D'Angelo and Canale}{2023}]{DAngelo2023}
D’Angelo, L., \& Canale, A. (2023). Efficient posterior sampling for Bayesian Poisson regression. \textit{Journal of Computational and Graphical Statistics}, \textbf{32}(3), 917-926.



\bibitem[\protect\citeauthoryear{Ferland et al.}{Ferland et al.}{2006}]{Ferland2006} 
Ferland, R., Latour, A., \& Oraichi, D. (2006). Integer-valued GARCH process. \textit{Journal of time series analysis}, \textbf{27}(6), 923-942.
\bibitem[\protect\citeauthoryear{Fokianos et al.}{Fokianos et al.}{2009}]{Fokianos2009} 
Fokianos, K., Rahbek, A., \& Tj\o stheim, D. (2009). Poisson autoregression. \textit{Journal of the American Statistical Association}, \textbf{104}(488), 1430-1439.
\bibitem[\protect\citeauthoryear{Fokianos and Tj\o stheim}{Fokianos and Tj\o stheim}{2011}]{Fokianos2011} 
Fokianos, K., \& Tj\o stheim, D. (2011). Log-linear Poisson autoregression. \textit{Journal of Multivariate Analysis}, \textbf{102}(3), 563-578.



\bibitem[\protect\citeauthoryear{Gamerman}{Gamerman}{1997}]{Gamerman1997} 
Gamerman, D. (1997). Sampling from the posterior distribution in generalized linear mixed models. \textit{Statistics and Computing}, \textbf{7}(1), 57-68.
\bibitem[\protect\citeauthoryear{Glynn et al.}{Glynn et al.}{2019}]{Glynn2019}
Glynn, C., Tokdar, S. T., Howard, B., \& Banks, D. L. (2019). Bayesian analysis of dynamic linear topic models. \textit{Bayesian Analysis}, \textbf{14}(1), 53-80.


\bibitem[\protect\citeauthoryear{Jiang et al.}{Jiang et al.}{2024}]{Jiang2024}
Jang, Y., Sundararajan, R. R., \& Barreto-Souza, W. (2024). A multivariate heavy-tailed integer-valued GARCH process with EM algorithm-based inference distribution in generalized linear mixed models. \textit{Statistics and Computing}, \textbf{34}(1), 56.
\bibitem[\protect\citeauthoryear{Jung et al.}{Jung et al.}{2006}]{Jung2006}
Jung, R. C., Kukuk, M., \& Liesenfeld, R. (2006). Time series of count data: modeling, estimation and diagnostics. \textit{Computational Statistics $\&$ Data Analysis}, \textbf{51}(4), 2350-2364.
\bibitem[\protect\citeauthoryear{Jung et al.}{Jung et al.}{2011}]{Jung2011}
Jung, R. C., Liesenfeld, R., \& Richard, J. F. (2011). Dynamic factor models for multivariate count data: An
application to stock-market trading activity. \textit{Journal of Business $\&$ Economic Statistics}, \textbf{29}, 73-85.


\bibitem[\protect\citeauthoryear{Keith et al.}{Keith et al.}{2008}]{Keith2008}
Keith, J. M., Kroese, D. P., \& Sofronov, G. Y. (2008). Adaptive independence samplers. \textit{Statistics and Computing}, \textbf{18}(4), 409-420.


\bibitem[\protect\citeauthoryear{Lee and Neal}{Lee and Neal}{2018}]{Lee2018}
Lee, C., \& Neal, P. (2018). Optimal scaling of the independence sampler: Theory and practice. \textit{Bernoulli}, \textbf{24}(3), 1636-1652.



\bibitem[\protect\citeauthoryear{Mengersen and Tweedie}{Mengersen and Tweedie}{1996}]{Mengersen1996}
Mengersen, K. L., \& Tweedie, R. L. (1996). Rates of convergence of the Hastings and Metropolis algorithms. \textit{The Annals of Statistics}, \textbf{24}(1), 101-121.



\bibitem[\protect\citeauthoryear{Pei and Zhu}{Pei and Zhu}{2024}]{Pei2024}
Pei, J., \& Zhu, F. (2024). Marginal likelihood estimation for the negative binomial INGARCH model. \textit{Communications in Statistics-Simulation and Computation}, \textbf{53}(4), 1814-1823.
\bibitem[\protect\citeauthoryear{Polson et al.}{Polson et al.}{2013}]{Polson2013}
Polson, N. G., Scott, J. G., \& Windle, J. (2013). Bayesian inference for logistic models using P\'{o}lya-Gamma latent variables. \textit{Journal of the American Statistical Association}, \textbf{108}(504), 1339-1349.
\bibitem[\protect\citeauthoryear{Polson et al.}{Polson et al.}{2019}]{Polson2019}
Polson, N. G., Scott, J. G., Windle, J., \&  Windle, M. J. (2019). Package ‘BayesLogit’.
\bibitem[\protect\citeauthoryear{Piancastelli et al.}{Piancastelli et al.}{2025}]{Piancastelli2025}
Piancastelli, L. S., \& Silva, R. B. (2025). Multivariate zero-inflated INGARCH models: Bayesian inference and composite likelihood approach. \textit{Statistics and Computing}, \textbf{35}(1), 19.


\bibitem[\protect\citeauthoryear{Reboredo et al.}{Reboredo et al.}{2023}]{Reboredo2023}
Reboredo, J. C., Barba-Queiruga, J. R., Ojea-Ferreiro, J., \& Reyes-Santias, F. (2023). Forecasting emergency department arrivals using INGARCH models. \textit{Health Economics Review}, \textbf{13}(1), 51.



\bibitem[\protect\citeauthoryear{Sim et al.}{Sim et al.}{2021}]{Sim2021}
Sim, T., Douc, R., \& Roueff, F. (2021). General-order observation-driven models: Ergodicity and consistency of the maximum likelihood estimator. \textit{Electronic Journal of Statistics}, \textbf{15}(1), 3349-3393.



\bibitem[\protect\citeauthoryear{Teerapabolarn}{Teerapabolarn}{2012}]{Teerapabolarn2012}
Teerapabolarn, K. (2012). The least upper bound on the Poisson-Negative Binomial relative error. \textit{Communications in Statistics-Theory and Methods}, \textbf{41}(10), 1833-1838.



\bibitem[\protect\citeauthoryear{Wei\ss~et al.}{Wei\ss~et al.}{2022}]{Weiss2022} 
Wei\ss, C. H., Zhu, F., \& Hoshiyar, A. (2022). Softplus INGARCH models. \textit{Statistica Sinica}, \textbf{32}(2), 1099-1120.

\bibitem[\protect\citeauthoryear{Wenzel et al.}{Wenzel et al.}{2019}]{Wenzel2019}
Wenzel, F., Galy-Fajou, T., Donner, C., Kloft, M., \& Opper, M. (2019). Efficient Gaussian process classification using P\'{o}lya-Gamma data augmentation. \textit{Proceedings of the AAAI Conference on Artificial Intelligence}, \textbf{33}(1), 5417-5424.



\bibitem[\protect\citeauthoryear{Xu et al.}{Xu et al.}{2020}]{Xu2020} 
Xu, X., Chen, Y., Chen, C. W., \& Lin, X. (2020). Adaptive log-linear zero-inflated generalized Poisson autoregressive model with applications to crime counts. \textit{The Annals of Applied Statistics}, \textbf{14}(3), 1493-1515.
\bibitem[\protect\citeauthoryear{Xu and Zhu}{Xu and Zhu}{2022}]{Xu2022} 

Xu, Y., \& Zhu, F. (2022). A new GJR-GARCH model for $Z$-valued time series. \textit{Journal of Time Series Analysis}, \textbf{43}(3), 490-500.

\bibitem[\protect\citeauthoryear{Xu and Zhu}{Xu and Zhu}{2026}]{Xu2026} 
Xu, Y., \&  Zhu, F. (2026). A zero-inflated Poisson asymmetric power GARCH model for $\mathbb{Z}$-valued time series. \textit{Communications in Mathematics and Statistics}, forthcoming. 



\bibitem[\protect\citeauthoryear{Yang et al.}{Yang et al.}{2022}]{Yang2022} 
Yang, K., Yu, X., Zhang, Q., \& Dong, X. (2022). On MCMC sampling in self-exciting integer-valued threshold time series models. \textit{Computational Statistics $\&$ Data Analysis}, \textbf{169}, 107410.
\bibitem[\protect\citeauthoryear{Yang et al.}{Yang et al.}{2023}]{Yang2023} 
Yang, K., Zhao, Y., Li, H., \& Wang, D. (2023). On bivariate threshold Poisson integer-valued autoregressive processes. \textit{Metrika}, \textbf{86}(8), 931-963.



\bibitem[\protect\citeauthoryear{Zhu et al.}{Zhu et al.}{2026}]{Zhu2026} 
Zhu, F., Xu, N., Li, Q., \& Ling, S. (2026). $\mathbb{Z}$-valued smooth transition GARCH models: Specification and testing. \textit{Journal of the American Statistical Association}, forthcoming. 

\end{thebibliography}
}
\end{document}